# Estimation of treatment effects following a sequential trial of multiple treatments

**John Whitehead*[+], Yasin Desai and Thomas Jaki**
*Medical and Pharmaceutical Research Unit*
*Department of Mathematics and Statistics*
*Lancaster University, Lancaster LA1 4YF, United Kingdom*
*j.whitehead@lancaster.ac.uk*

*14 December 2018*

**Abstract**

When a clinical trial is subject to a series of interim analyses as a result of which the study may be terminated or modified, final frequentist analyses need to take account of the design used. Failure to do so may result in overstated levels of significance, biased effect estimates and confidence intervals with inadequate coverage probabilities. A wide variety of valid methods of frequentist analysis have been devised for sequential designs comparing a single experimental treatment with a single control treatment. It is less clear how to perform the final analysis of a sequential or adaptive design applied in a more complex setting, for example to determine which treatment or set of treatments amongst several candidates should be recommended.

This paper has been motivated by consideration of a trial in which four treatments for sepsis are to be compared, with interim analyses allowing the dropping of treatments or termination of the trial to declare a single winner or to conclude that there is little difference between the treatments that remain. The approach taken is based on the method of Rao-Blackwellisation which enhances the accuracy of unbiased estimates available from the first interim analysis by taking their conditional expectations given final sufficient statistics. Analytic approaches to determine such expectations are difficult and specific to the details of the design, and instead "reverse simulations" are conducted to construct replicate realisations of the first interim analysis from the final test statistics. The method also provides approximate confidence intervals for the differences between treatments.

**Key words:** *Adaptive designs, Estimating treatment effects, Multiple treatment trials, Rao-Blackwellisation, Sequential designs.*

**Correspondence to John Whitehead, 6 Alderman Road, Lancaster, LA1 5FW, UK.*
[+]*E-mail: j.whitehead@lancaster.ac.uk*

## 1. Introduction

This paper is motivated by a design for a trial in sepsis [1], with the objective of comparing four treatments for sepsis in respect of survival of the patient to 28 days, analysed as a binary response and referred to as "success". Although one of the treatments is standard care, it is given no special privileges: all four treatments are dealt with in the same way. At each of a series of interim analyses, all pairwise comparisons of remaining treatments are made. Any treatment that is worse than any other according to pre-determined criteria is eliminated from the trial. If only one treatment remains, it is declared the winner and the trial stops. If all remaining treatments are sufficiently similar to one another, they are declared joint winners and the trial stops. Otherwise the trial continues to the next interim analysis. At the end of such a trial, how should the differences between the performances of the treatments be estimated, bearing in mind the potential biases introduced by the elimination and stopping rules [2]?

Estimation following a sequential trial has a rich statistical literature. For the case of trials comparing two treatments, methods may be based on orderings of the final sample space [3-6] or on direct adjustment of the maximum likelihood estimate to reduce its bias [7]. Overviews of alternative approaches are available [8, 9]. For adaptive designs, which are not entirely pre-defined, estimation methods can be devised by extending the ordering approach [10, 11] or through shrinkage approaches [12, 13]. In this paper, an approach based on Rao-Blackwellisation [14, 15] will be developed. This is a method that has been adopted by previous authors for certain specific designs [16, 17].

It is difficult to generalise approaches based on orderings of the final sample space to trials of multiple treatments because there are so many possible final samples and it becomes unclear how to determine which provide stronger evidence favouring a given alternative over the null than the sample observed. Direct adjustment of maximum likelihood estimates depends on knowledge of the distribution of the final sample statistics around the stopping boundary. While this can be characterised for a comparison of two treatments that relies on a single test statistic, it is far more challenging to achieve for multiple treatments compared in respect of several pairwise test statistics. Here the Rao-Blackwellisation approach will be developed. This is based on the expected value of an unbiased estimate computed at the first interim analysis (and thus unaffected by any stopping rules), conditional on sufficient statistics computed at the end of the study. Rather than finding this test statistic and its standard error analytically, it will be determined by reverse simulation. That is, starting with the final values of the numbers of patients and the numbers of successes for each treatment (and when present, within each stratum), hypergeometric sampling will be used to create possible samples at each earlier interim analysis until that at the first interim has been recreated. Only those sequences that are consistent with continuation to the observed end of the sequential procedure are accepted. The mean and variance of unbiased estimates from each acceptable replicate simulated first interim analysis are then used to provide unbiased estimates and approximate confidence intervals that allow for the sequential nature of the design.

The approach developed has the potential for implementation following a wide range of multiple treatment trials and flexible adaptive designs. It is often easier to work backwards from the end of the trial and determine which sequences of data would have led to continuation to the final sample, than to project such sequences from the outset. Much of the development and evaluation of the method will be made in the context of a conventional sequential comparison of just two treatments because that setting is simpler analytically and computationally, and because it allows comparisons to be made with more established methods.

In the next section the trial design described in [1] is reviewed. Instead of using the elimination and stopping rules proposed in that paper, the performance of alternative rules based on the triangular test [8] is examined. It must be emphasised that this design serves here only as an illustration of the new estimation approach. The latter could be applied to a wide range of multiple treatment designs and indeed other forms of flexible adaptive design. In Section 3, the simple comparative triangular design that forms the basis of the four treatment evaluation is extracted and examined in isolation, and in Section 4, naïve and orderings-based analyses are developed in the two-treatment context, together with two forms of the Rao-Blackwellisation approach. The new method is applied to simulated data from the four treatment design in Section 5, and Section 6 draws together conclusions from this work.

## 2. A design for the comparison of four treatments

The design introduced by Magaret et el. [1] comprised up to four successive analyses based on constant nominal $\alpha$-levels. Here the overall structure of that design is retained but a different approach is taken to the elimination and stopping rules: one which will allow the extraction of a simple triangular test [8] for examination in the central sections of this paper.

Treatment effects are expressed in terms of odds ratios for success. The value 1.5 is taken to be of clinical importance. The probability of success (survival to 28 days) for a patient receiving treatment $T_i$ is denoted by $p_i$, i = 1, ..., 4. The log-odds ratio for treatment $T_i$ relative to $T_j$ is denoted by $\theta_{ij} = \log[\{p_i(1 - p_j)\}/\{p_j(1 - p_i)\}]$. The design seeks to satisfy the following requirements. *Type I error requirement:* For any treatment $T_i$, if there is another treatment $T_j$, $i \neq j$ such that $p_i = p_j$, then the probability that the trial finds $T_i$ to be the sole winner is to be $\leq 0.025$. *Power requirement:* For any pair of treatments $T_i$ and $T_j$, if $T_i$ is superior to $T_j$ to the extent that $\theta_{ij} = \log(1.5)$, then the probability that $T_j$ would be eliminated from the study is to be $\geq 0.90$.

Interim analyses occur whenever 36 new patient outcomes become available on each of the treatments remaining in the study. The maximum sample size is set at 2772 patient responses and interim analyses continue until the trial stopping rules are satisfied or else it is impossible to assign 36 more patients to all remaining treatments within this quota. The probability that not all treatment comparisons will be resolved after 2772 patient responses have been observed is small. If all four treatments were to remain in the trial, the maximum number of interim analyses would be about 20: more could occur if treatments were eliminated.

At the $k^{th}$ interim analysis, every pair of treatments $T_i$ and $T_j$ will be compared in terms of the statistics $Z_{ijk}$ and $V_{ijk}$ where

$$Z_{ijk} = \frac{n_{jk}S_{ik} - n_{ik}S_{jk}}{n_{ik} + n_{jk}} \quad \text{and} \quad V_{ijk} = \frac{n_{ik}n_{jk}\left(S_{ik} + S_{jk}\right)\left(n_{ik} + n_{jk} - S_{ik} - S_{jk}\right)}{\left(n_{ik} + n_{jk}\right)^3}. \qquad (1)$$

Here $n_{ik}$ denotes the number of patient responses available for patients on Treatment $T_i$, and $S_{ik}$ the number of those who have survived to Day 28. For a stratified version of the procedure, the statistics shown are computed separately within each stratum, and then summed over strata to provide the values of $Z_{ijk}$ and $V_{ijk}$ to be used.

At the $k^{th}$ interim analysis, it is concluded that $T_i$ is better than $T_j$ if $Z_{ijk} \geq 10.90266 + 0.12380V_{ijk}$, no different from $T_j$ if $Z_{ijk} \in (10.90266 - 0.37140V_{ijk}, -10.90266 + 0.37140V_{ijk})$, and worse than $T_j$ if $Z_{ijk} \leq -10.90266 - 0.12380V_{ijk}$. If the interval used to judge no difference is empty because the left-hand limit is larger than the right-hand limit, then the no difference conclusion is not possible. Whenever one treatment is found to be worse than another according to this criterion, that treatment is eliminated from the trial. Randomisation continues between the remaining treatments, and interim analyses continue to take place whenever 36 new outcomes have become available for each remaining treatment. The trial stops when only one treatment remains, or when all remaining treatments are found to be no different from one another. For the purposes of the simulations conducted here, the trial also stops if a further interim analysis would require the total number of patients to exceed 2772, although in practice investigators might choose an alternative strategy as discussed later in this section.

The elimination and stopping rules, as they relate to a comparison between one pair of treatments, are shown in Figure 1. Each interim analysis is represented by a disc on the boundaries, and at the $k^{th}$ interim analysis, the value of $Z_{ijk}$ is plotted against that of $V_{ijk}$, and the conclusion indicated is drawn. The design has been developed from a double triangular design devised to compare two experimental treatments [8, 18]. The boundaries are computed to satisfy the type I error and power requirements mentioned above, interpreted for the simple case of two treatments. Computation is based on the SEQ function of SAS, following [19, 20], but using the four boundary option of SEQ. The increment in information between interim analyses for this double triangular test is V = 4.40337. When $p_1 = 0.40$ and $p_2 = 0.50$ or when $p_1 = 0.50$ and $p_2 = 0.60$ (both corresponding to an odds ratio of 1.5), this corresponds to an increase in sample size between interim analyses of 35.58 per treatment, which is rounded up to 36 in this application.

Applied to the case of four treatments, the type I error and power requirements specified at the beginning of this section are valid. The probability that $T_1$ is declared the sole winner, when in fact $p_1 = p_2$ is greatest when the success rates on $T_3$ and $T_4$ are both zero so that there is no chance of them being declared either sole or joint winners. There would also be a negligible chance that they would be declared no different from $T_1$ or from $T_2$ or from both. In this circumstance, the probability that $T_1$ would be declared the sole winner is therefore equal to the probability of $T_1$ being found better than $T_2$ in the double triangular test when $\theta_{12} = 0$: that is 0.025. Furthermore, the probability that $T_1$ is eliminated, when $\theta_{12} = \log(1.5)$ is least when the success rates on $T_3$ and $T_4$ are both zero so that there is no chance that $T_1$ would be eliminated relative to them. In this circumstance, the probability that $T_1$ would be eliminated is therefore equal to the probability of $T_2$ being found better than $T_1$ in the double triangular test when $\theta_{12} = \log(1.5)$: that is 0.900.

Properties of the design estimated from million-fold simulations, are shown in Table 1 below. In each of the Cases 1-12, one set of treatments share a high success rate and the rest share a low rate, with the odds ratio between the two rates being 1.5. In Cases 13-16, all success rates are equal. Also shown are "Mixed Cases". For these, we imagine that the trial is conducted at four centres each recruiting equal numbers of patients. In the simulations, the 36 patients recruited to each treatment for each new interim analysis are distributed amongst the centres at random. The four centres in the mixed cases each have a different set of success probabilities, namely the four sets shown in the cases above. In the simulations for the mixed cases, the statistics Z and V given in (1) are stratified for centre: that is the four within-centre values of Z and V are calculated and then summed to provide the values to be compared with the stopping boundaries.

In Cases 1-4 and Mixed Case I, the probability that $T_4$ is correctly eliminated exceeds 0.90, as specified in the power requirement. This is true for $T_2$ and $T_3$ as well, although these results are not shown: in general the full results reflect the symmetry of each scenario. Treatment $T_1$ is correctly selected with a probability exceeding 0.80: this is a desirable feature, although not part of the formal specification. In Cases 5-8 and Mixed Case II, the probability of wrongly declaring $T_1$ to be the winner is no more than 0.026, (essentially) satisfying the type I error requirement. The probability of eliminating $T_4$ is well above the value of 0.90 of the power requirement. The probability of correctly declaring $T_1$ and $T_2$ to be joint winners is above 0.90, except for Case 8 where it is 0.885. In Cases 9-12 and Mixed Case III, the probability that $T_1$ wins is 0.005 and the probability that $T_4$ is eliminated is greater than 0.975. The probability of correctly identifying the three joint winners is greater than 0.814. Finally, in Cases 13-16 and Mixed Case IV, the probability that $T_1$ wins is 0.002 or less in all cases. The probability of correctly identifying all four treatments as no different is greater than 0.748, except for case 16 where it is 0.591.

Average total sample sizes at termination are around 1400-2400. Sample sizes are smaller when success probabilities are close to ½, and larger when they are close to 1 or to 0. They are also smaller when there is a single treatment that is more efficacious than the others, or when there are two good treatments. Cases where three or all four treatments are equally efficacious require larger sample sizes before a conclusion is reached. Ethically, this is sound, as if all treatments are the same, no group of patients is being disadvantaged by being in the trial. By the same token, patients not in the trial are under no disadvantage due to the length of the trial. The full results show that sample sizes on poor treatments tend to be small, those on good treatments to be large, indicating the effectiveness of eliminating poor treatments. The percentage of inconclusive trials was 26.6% in Case 16. In all other cases, such percentages are small or negligible. If the trial ends without identifying a single winner or concluding that there is no difference between the remaining treatments, then investigators can accept the result available, or else recruit additional patients to force a conclusion.

The construction of the decision rules of the design guarantees that it is not possible to declare two treatments to be no different from one another during the first 6 interim analyses (see Figure 1). It is possible to stop at any analysis to conclude that one of the treatments is better than all of the others, but the evidence has to be very clear. Minimal evidence for $T_2$ to be eliminated relative to $T_1$ at the first interim analysis requires 23 successes out of 36 on $T_1$ and none on $T_2$: the corresponding one-sided nominal p lies well below 0.00001. In fact, under most realistic scenarios, the probability of stopping at one of first three interim analyses is negligible.

It can be seen that the procedure presented here achieves the type I error and power requirements specified, and has other desirable properties in terms of high probabilities of appropriate conclusions and relatively low expected sample sizes. It must be stressed that these interim analyses are very simple to carry out. The following information on all patients randomised 28 days ago or earlier is all that is needed: Patient identification number; Treatment centre and any other baseline stratification factors; Date of randomisation; Treatment arm ($T_1$, $T_2$, $T_3$ or $T_4$); and Survived to Day 28 (YES or NO). All but the last are available for a month before the patient is to be included in interim analyses. More extensive reviews of the data might be planned, perhaps to coincide with every 4th or every 5th interim analysis. Interim analyses are to be conducted whenever the average number of patient responses per remaining treatment collected since the previous interim analysis reaches 36. Ideally, this should be 36 patients per treatment, but the formulae given at (1) can be used when sample sizes are unequal, and the accuracy will remain good provided that sample sizes per treatment are approximately equal. The method is also likely to be forgiving of slight slippage from an average of exactly 36 new patients per treatment.

## 3. Conventional post-trial estimation for a simple triangular test

Now consider a comparison between just two treatments, $T_1$ and $T_2$. A series of up to 20 interim analyses are conducted, at the $k^{th}$ of which the statistics $Z_{12k}$ and $V_{12k}$ defined in (1) will be computed. Here, they will be denoted simply as $Z_k$ and $V_k$, and the log-odds ratio $\theta_{12}$ by $\theta$. The trial will be stopped with the conclusion that $T_1$ is better than $T_2$ if $Z_k \geq 10.93898 + 0.123134V_k$, or with the conclusion that $T_1$ is no better than $T_2$ if $Z_k \leq -10.93898 + 0.369402V_k$. The design is constructed using published code [19, 20], and the risk of one-sided type I error is set to 0.025 and the power for an odds ratio of 1.5 to 0.90. Note that the boundaries differ slightly from those used in the four treatment case, because the latter were based on the properties of pairwise double triangular tests. Here, $T_1$ can be thought of as the experimental treatment and $T_2$ as the control: the design is asymmetric in dealing with the treatments. The maximum value of V is $V_{20} = 88.8380$, at which point the stopping boundaries meet. Hence $V_1 = 4.4419$. For $p_1 = 0.60$ and $p_2 = 0.50$, so that $\theta = \log(1.5)$, the total sample size per interim analysis is approximately 72 (36 per treatment). In simulations reported here, additional interim analyses are conducted beyond the 20 initially planned, up to a maximum of 25, if increments in V fall short of the anticipated value of 4.4419 and no boundary has been reached. In practice, if increments in V are observed to be low, then sample sizes per interim can be increased.

Table 2 shows the results of 12 simulated realisations of this triangular design, ordered by increasing strength of evidence that $T_1$ is better than $T_2$. Also given are results of a naïve analysis in which the sequential nature of the design is ignored, and a valid analysis based on the ordering of Fairbanks and Madsen [21]. For the naïve analysis, the estimated value of $\theta$ is taken to be $\hat{\theta} = Z^*/V^*$ with standard error se = $1/\sqrt{V^*}$, and the corresponding 95% confidence interval ($\theta_L$, $\theta_U$) is ($\hat{\theta} \pm 1.96$se), where $Z^*$ and $V^*$ are the values of Z and V found from the final dataset. The orderings analysis is computed following [19] and [20]. In each computation the value of $V_i$ is taken to be equal to $i^*V_{20}/20$. In practice the true values of the $V_i$ would be used, but the approximation is used here for simplicity, and to allow readers to check the computation of the estimates. The analysis methods developed in the next section do not depend on the intermediate values of the $V_i$. The bias-adjusted estimate [7] has no corresponding accurate method for computing confidence intervals and for that reason, it is not explored here.

The orderings analysis provides valid p-values and reduces estimates of $\theta$ when the upper boundary is crossed and increases them in the case of the lower boundary. It provides totally satisfactory results based on the actual sequential design used. However, it is difficult to see how it might be generalised for use following a sequential comparison of more than two treatments.

## 4. Post-trial estimation based on Rao-Blackwellisation for a simple triangular test

The Rao-Blackwellisation approach [14, 15] is based on the estimate $\hat{\theta}_1 = Z_1 / V_1$ deduced from the data available at the first interim analysis, which is unbiased for $\theta$ as it does not depend on the stopping rule in any way. Consequently, the estimate $\tilde{\theta} = E\left(\hat{\theta}_1 \middle| Z^*, V^*\right)$, is also unbiased for $\theta$ and has smaller variance. The estimate is truncation-adaptable, meaning that it depends only on the form of the interim analyses that were performed and not on those that were planned to take place but did not. Orderings analyses are also truncation-adaptable, but the bias adjusted method [7] is not. The estimator $\tilde{\theta}$ achieves minimum variance within the class of unbiased truncation-adaptable estimators [22].

Now $E\left(\tilde{\theta}\right) = \theta$, and

$$\mathrm{var}\left(\tilde{\theta}\right) = \mathrm{var}\left\{E\left(\hat{\theta}_1 \middle| Z^*, V^*\right)\right\} = \mathrm{var}\left(\hat{\theta}_1\right) - E\left\{\mathrm{var}\left(\hat{\theta}_1 \middle| Z^*, V^*\right)\right\} = \left(1/V_1\right) - E\left\{\mathrm{var}\left(\hat{\theta}_1 \middle| Z^*, V^*\right)\right\}.$$

In order to compute confidence intervals, it will be assumed that the pivot $\left\{\tilde{\theta} - E\left(\tilde{\theta}\right)\right\} \Big/ \sqrt{\mathrm{var}\left(\tilde{\theta}\right)}$ follows the standard normal distribution and that $E\left\{\mathrm{var}\left(\hat{\theta}_1 \middle| Z^*, V^*\right)\right\}$ can be reliably estimated by $\mathrm{var}\left(\hat{\theta}_1 \middle| Z^*, V^*\right)$. Thus the standard error of $\tilde{\theta}$ is given by

$$\mathrm{se}\left(\tilde{\theta}\right) = \sqrt{\left\{\left(1/V_1\right) - \mathrm{var}\left(\hat{\theta}_1 \middle| Z^*, V^*\right)\right\}}\ , \qquad (2)$$

and an approximate 95% confidence interval for $\theta$ is $\left(\tilde{\theta} \pm 1.96\mathrm{se}\left(\tilde{\theta}\right)\right)$. It is unlikely that either of the assumptions on which this approach is based are more than approximately true. The accuracy of the derived confidence intervals should be evaluated by simulation for any given application. The theoretical basis for the unbiasedness of the estimate $\tilde{\theta}$ is far stronger than that for the accuracy of the confidence interval.

Two methods for evaluating $\tilde{\theta}$ and $\mathrm{se}\left(\tilde{\theta}\right)$ will now be developed. The first, Method RB1, is an analytical approach depending on known properties of the triangular test. It is infeasible to generalise RB1 to the four treatment case, and it is included here for comparison and checking. Method RB2 employs reverse simulation to re-create replicate observations of $Z_1$ and $V_1$, and is applicable in complicated situations such as a comparison of four treatments.

### *4.1 Method RB1*

Denote the lower and upper stopping limits for $Z_k$ at the $k^{th}$ interim analysis by $\ell_k$ and $u_k$ respectively, k = 1, 2, ... . The sequential design based on the first n of these interim analyses, which is then truncated, is denoted by $R_n$. The interim analysis at which the design $R_n$ actually stops will be denoted by $K_{[n]}$, and the corresponding final values of the test statistics by $Z_{[n]}$ and $V_{[n]}$. Equation (5.38) of [8] defines the function $f_{[n]}(z, k, \theta)$ to be

$$f_{[n]}(z,k,\theta) = \lim_{\delta z\to 0}\frac{1}{\delta z}P\left(Z_{[n]}\in(z,z+\delta z),K_{[n]}=k\right),\quad k = 1, ..., n. \tag{3}$$

The sequence of functions $f_{[i]}(z, k, \theta)$ for $z < \ell_k$ or $z > u_k$, k = 1, ..., n − 1, and $f_{[n]}(z, n, \theta)$, together provide a density for the final position of the test statistics ($Z_{[n]}$, $V_{[n]}$) over all of their possible final values. Using Equation (5.41) of [8], it can be shown that for any $\theta$, $f_{[n]}(z_n,n,\theta) = \exp\left(z_n\theta - \theta^2 V_n\right) f_{[n]}(z_n,n,0)$, as pointed out in [15].

Now, let

$$F_{[n]}(z_n,n,\theta) = \int_{-\infty}^{z_n} f_{[n]}(s,n,\theta)\,ds.$$

This is the probability that the design $R_n$ stops at the $n^{th}$ interim analysis with $Z_{[n]} \le z_n$. In fact, it is the probability that any design which shares with $R_n$ the stopping limits for its first n interim analyses stops at the $n^{th}$ interim analysis with $Z_{[n]} \le z_n$. This function can be evaluated using the SAS function SEQ. Note that

$$F_{[n]}(z_n,n,\theta) = \exp\left(z_n\theta - \theta^2 V_n\right) F_{[n]}(z_n,n,0). \tag{4}$$

The probability that the design $R_n$ stops at the $n^{th}$ interim analysis with $Z_{[n]} \in (z_n - \delta z, z_n + \delta z)$ is given by $F_{[n]}(z_n+\delta z,n,\theta) - F_{[n]}(z_n-\delta z,n,\theta)$.

Next, consider the adjusted sequential test, in which the first lower limit is amended to be $\ell_1 + t$, $t \in (0, u_1 - \ell_1)$. The functions corresponding to $f_{[n]}$ and $F_{[n]}$ for this design will be denoted by $f_{[n]}^{(t)}$ and $F_{[n]}^{(t)}$ respectively. Then, the probability that the sequential design $R_n$ starts with a value of $z_1$ lying above $\ell_1 + t$ and then later stops at the $n^{th}$ interim analysis with $Z_{[n]} \in (z_n - \delta z, z_n + \delta z)$ is given by $F_{[n]}^{(t)}(z_n+\delta z,n,\theta) - F_{[n]}^{(t)}(z_n-\delta z,n,\theta)$. Denote the conditional probability that $Z_1$ lies above $\ell_1 + t$, given that the trial stops at the $n^{th}$ interim analysis with $Z_{[n]} \in (z_n - \delta z, z_n + \delta z)$ by $S(t; \theta)$. Then $S(t;\theta) = P\left(Z_1 > \ell_1 + t \,\middle|\, K_{[n]} = n, Z_{[n]} \in (z_n - \delta z, z_n + \delta z)\right)$. It follows, using (4), that

$$S(t) = \frac{F_{[n]}^{(t)}(z_n+\delta z,n,0) - F_{[n]}^{(t)}(z_n-\delta z,n,0)}{F_{[n]}(z_n+\delta z,n,0) - F_{[n]}(z_n-\delta z,n,0)}, \tag{5}$$

where the value of $\theta$ is suppressed in the notation for S. This confirms the sufficiency of the statistics $Z_{[n]}$ and $J_{[n]}$. It can be shown that

$$\int_0^{u_1-\ell_1} S(t)dt = E\left(Z_1-\ell_1 \middle| K_{[n]}, Z_{[n]}\right) \text{ and } 2\int_0^{u_1-\ell_1} tS(t)dt = E\left\{\left(Z_1-\ell_1\right)^2 \middle| K_{[n]}, Z_{[n]}\right\}. \quad (6)$$

Suppose that a trial with stopping limits at the first n interim analyses of ($\ell_1$, $u_1$), ... , ($\ell_n$, $u_n$) stops with $Z_n = z_n$. Thus $K_{[n]} = n$ and $Z_{[n]} = z_n$. The value of $F_{[n]}(z_n+\delta z,n,0)-F_{[n]}(z_n-\delta z,n,0)$ can be evaluated using the SAS function SEQ for the trial stopping limits with the modification that the n^th^ continuation region is ($z_n - \delta z$, $z_n + \delta z$). The value of $\delta z$ is chosen to be small, but large enough for the resulting n^th^ continuation probability to be reported with a reasonable number of decimal places. Then $F_{[n]}^{(t)}(z_n+\delta z,n,0)-F_{[n]}^{(t)}(z_n-\delta z,n,0)$ is evaluated in a similar way, but for a design with first continuation region given by ($\ell_1$ + t, $u_1$) for a grid of values of t between 0 and $u_1 - \ell_1$. This allows S(t) to be found from (5) for the same grid of values, from which the conditional mean and standard deviation of $Z_1$ given $K_{[n]} = n$ and $Z_{[n]} = z_n$ can be found from (6) using numerical integration. Note that the function SEQ is constructed for stopping limits for $Z_n/\sqrt{V_1}$. This necessitates some intricate programming in order to obtain the correct answers.

### *4.2 Method RB2*

For reverse simulation, the estimate $\tilde{\theta} = E\left(\hat{\theta}_1 \middle| \mathbf{S}^*, \mathbf{n}^*\right)$ is used, where **S*** and **n*** are the vectors of numbers of successes and numbers of patients, by treatment, in the final dataset. The final interim analysis will be taken to be the $K^{th}$. The number of successes on $T_i$ at the $k^{th}$ interim analysis, $S_{ik}$, is simulated as a hypergeometric observation, being the number of successes in a draw of $n_{ik}$ patient responses from a total of $n_{i,k+1}$ responses of which $S_{i,k+1}$ are successes, i = 1, 2; k = K – 1, K – 2, ..., 1. For each replicate simulation, the estimate $\hat{\theta}_1 = Z_1/V_1$ is found from (1) using the simulated numbers of successes on the two treatments at the first interim analysis. All values of $Z_k$ and $V_k$, k = 1, ..., K – 1, are checked, and any simulated sample path that corresponds to a trial that would have stopped according to the sequential design is deleted from the set of simulated runs. Then the mean and variance of the remaining values of $\hat{\theta}_1$ are used as $\tilde{\theta}$ and $\operatorname{var}\left(\hat{\theta}_1 \middle| \mathbf{S}^*, \mathbf{n}^*\right)$ respectively. The latter is used in place of $\operatorname{var}\left(\hat{\theta}_1 \middle| Z^*, V^*\right)$ in (2) to provide a value for $se\left(\tilde{\theta}\right)$. The set of simulated realisations of the first interim analysis can be used in a similar way to provide unbiased estimates of the success probabilities $p_1$ and $p_2$, allowing for the sequential nature of the trial.

### *4.3 Evaluation of Methods RB1 and RB2*

Table 3 presents results from analyses of the 12 cases presented in Table 2 using Rao-Blackwellisation methods. For Method RB1, the value of $\delta z$ in (5) was set at 0.01, and a grid of 100 points was used to evaluate the integrals in (6). For Method RB2, 10 million-fold reverse simulations were generated, and the first column in the RB2 section of the table shows that between 17.0% and 99.3% of the replicates were complete: that is they corresponded to sequential trials that would not have stopped prior to the observed final interim analysis. Figure 2 shows the estimate and confidence intervals from the Rao-Blackwellisation approaches and the same quantities from the naïve and orderings analysis, plotted against the values of the naïve estimates. The value of the naïve estimate is subtracted from all quantities, in order to provide a clearer view of the differences between the methods. The 12 cases are ordered with respect to the naïve estimates, and so Cases 1 to 12 are the points indicated by crosses running from left to right. The vertical line at $\theta$ = 0.2462 represents the value of treatment effect at which the trend of the plot of Z against V would head for the tip of the triangle, for this is the average of the boundary slopes.

When the naïve estimate lies below 0.2462, both adjustments increase the magnitude of the estimate, with those due to the Rao-Blackwell estimate being greater. When the naïve estimate lies above 0.2462, both adjustments reduce the magnitude of the estimate, with those due to the Rao-Blackwell estimate again being greater. Adjustments using Method RB1 are a little more extreme than those using RB2.

The naïve confidence limits are narrowest and will fail to meet the target coverage probability. The Method RB1 leads to the widest intervals, followed closely by RB2. When the naïve estimate lies below 0.2462, adjusted limits lie above the corresponding naïve limits and when it lies below 0.2462 they lie below. This effect is greatest for the Rao-Blackwell approaches. In cases in which there is a large overshoot of the boundary at the final interim analysis (Cases 6, 8, 9 and 11), the adjustments for sequential analysis have the greatest effect on the estimate of $\theta$. In cases where the overshoot is small (Cases 5, 10 and 12), the adjustments for sequential analysis have less effect on the estimate of $\theta$. The SAS programs leading to the RB1 and RB2 analyses shown in Table 3 are provided as supplementary material of this paper.

Table 4 presents the results of 1000-fold simulations of the naïve approach and of Methods RB1 and RB2 for three true values of $\theta$. These are the null value, 0; the alternative value log(1.5) = 0.405; and between these the value 0.246 which is the average of the two boundary slopes. In each case the control success probability was set at $p_C$ = 0.6. The results from the naïve approach confirm that there is a problem to be addressed. In particular, when $\theta$ = 0.405 the effect of the treatment is systematically overestimated, and in all three cases the coverage probability of confidence intervals is inadequate. For Method RB1, the grid is again constructed of 100 points and the value of $\delta z$ set as 0.01. For Method RB2, for reasons of computing time, one million replicate reverse simulations were used rather than the ten million that underlay the results presented in Table 3. The Rao-Blackwellised estimate $\tilde{\theta}$ is presented with its standard deviation computed from the 1000 replicate values and its standard error, which is the mean of the values computed from (2). These two values are close to one another, in support of the basis of computation. For both methods, the bias of $\tilde{\theta}$ is small. The coverage probabilities of the 95% confidence intervals are around 0.970 (and significantly greater than 0.95 at the one-sided 2.5% level) in each case. They could therefore be used in practice as conservative computations.

## 5. Application to the comparison of four treatments

A single set of simulated data consistent with the design proposed in Section 1 is used to illustrate the implementation of Rao-Blackwellisation in a more complicated situation. Table 5 displays the data from this single realisation. This summary is sufficient for analysis according to Method RB2. There are six pairwise treatment comparisons to consider. Table 6 presents the final values of the test statistics Z and V for each of these comparisons. Treatment $T_2$ was eliminated in comparison with $T_1$ at the $4^{th}$ interim analysis, and $T_4$ followed at the $5^{th}$. This left $T_1$ and $T_3$, which continued to be monitored until the $12^{th}$ interim analysis, at which point $T_1$ was found to be the winner.

Whether the analysis is conducted allowing for the sequential design used or not, two options are available for the final analysis. Option 1 is to use all data available on each treatment in making each comparison. Option 2 is to restrict the data used in any pairwise comparison to that collected from patients randomised when both treatments were still in contention. This is the form of analysis reflected in the values of Z and V displayed in Table 6. It avoids biases that may be caused by any temporal effects on the nature of the patients recruited, on the manner in which treatments were administered, or on how observations were recorded. Option 2 will be adopted here.

To implement Option 2, three separate reverse simulations have to be performed. To compute the estimate $\tilde{\theta}_{13}$ and its standard error, reverse simulation is conducted from the $12^{th}$ interim analysis, at which $T_1$ was found to be better than $T_3$, leading to the termination of the whole trial. From Table 5, it can be seen that at the $12^{th}$ interim analysis at Centre 1, $T_1$ had been administered to 103 patients with 83 successes and $T_3$ to 111 patients with 85 successes. At the $11^{th}$ interim analysis at Centre 1, $T_1$ had been administered to 98 patients and $T_3$ to 102 patients. For the reverse simulation, the number of successes on $T_1$ at Centre 1 is generated as a hypergeometric random variable: the number of successes from 98 patients drawn randomly from 103 of which a total of 83 are successes. The number of successes on $T_3$ at Centre 1 is generated similarly, as are the success counts for other centres. These success counts are then used to generate the numbers of successes on the two treatments at the $10^{th}$ interim analysis, and so on back to the first interim analysis. In the reverse simulation, the numbers of patients and of successes on $T_4$ at the $5^{th}$ interim analysis is taken to be as recorded in Table 5, and the numbers of successes at earlier interim analyses are filled in by hypergeometric simulation; for $T_2$ the reverse simulation begins at the $4^{th}$ interim analysis.

The next step is to determine which of the reverse simulated runs are consistent with the outcome of the trial, and to delete those which are not. For each reverse simulated run, every remaining treatment comparison is considered at each interim analysis in turn. The relevant stratified values of Z and V can be computed from the simulated success counts. Consider the comparison between treatments $T_i$ and $T_j$, $i \neq j = 1, 2, 3, 4$. First, consider an interim analysis which in the real trial is the last for both $T_i$ and $T_j$. In such a case the reverse simulated data for both treatments will be identical to those used in the actual trial, and the conclusions will be the same. No runs will be deleted on the basis of these data.

Now consider an interim analysis which in the real trial is the last for $T_i$, but after which $T_j$ continued to be observed. If, in the real trial $T_j$ was found better than $T_i$ at this interim analysis, then any reverse simulation for which this did not occur is deleted. It is possible that in the real trial $T_j$ was not found better than $T_i$ at this interim analysis, $T_i$ being eliminated in comparison with another treatment. In this case, any reverse simulation in which $T_i$ was found to be better than, or worse than, $T_j$ is eliminated. Furthermore, any reverse simulation run that ends at this interim analysis with the conclusion that there is no difference between any of the remaining treatments will be deleted.

Finally, consider an interim analysis which in the real trial is not the last for either $T_i$ or $T_j$. Any reverse simulation for which $T_j$ was found better than, or worse than, $T_i$ at this interim analysis is deleted. Once more, any reverse simulation run that ends at this interim analysis with the conclusion that there is no difference between any of the remaining treatments will be deleted.

For each of the reverse simulation runs that remains after the deletion process, the estimate $\hat{\theta}_{13} = Z_{131}/V_{131}$ is found from (1) using the reverse simulated stratified test statistics for the comparison of $T_1$ and $T_3$ from the first interim analysis. The mean of the values of $\hat{\theta}_{13}$ provides the RB2 estimate $\tilde{\theta}_{13}$ and the corresponding variance provides $\text{var}\left(\hat{\theta}_{13} \middle| \mathbf{S^*}, \mathbf{n^*}\right)$. The latter is used in a suitably amended version of equation (2) to provide a value for $\text{se}\left(\tilde{\theta}_{13}\right)$.

A second reverse simulation is then run, starting at the 5th interim analysis, and using the actual numbers of successes on $T_1$, $T_3$ and $T_4$ at each centre at that analysis as the starting point for each reverse simulation. Following the deletion of runs that would have been incomplete, $\tilde{\theta}_{14}$ and $\tilde{\theta}_{34}$ and their corresponding standard errors are found. The third reverse simulation starts at the 4th interim analysis and uses the actual numbers of successes observed on all treatments at each centre at that analysis as the starting point for each reverse simulation. This provides the estimates $\tilde{\theta}_{12}$, $\tilde{\theta}_{23}$ and $\tilde{\theta}_{24}$ and their corresponding standard errors.

In the results that follow, one modification of the method implemented in the unstratified case is made. For the purposes of computing the $\tilde{\theta}_{ij}$ and their standard errors only, $V_{ijc1}$ is replaced by $V'_{ijc1}$, where

$$V'_{ijc1} = \frac{n_{ic1} n_{jc1} \left(S_{ic1} + S_{jc1}\right)\left(n_{ic1} + n_{jc1} - S_{ic1} - S_{jc1}\right)}{\left(n_{ic1} + n_{jc1}\right)^2 \left(n_{ic1} + n_{jc1} - 1\right)}, \tag{7}$$

and the additional subscript c indicates the centre, c = 1, ..., 4. The usual expression for $V_{ijc1}$, is used during the conduct of the trial and when assessing whether simulated trial runs are complete. However, it is $\hat{\theta}_{ij1} = \left(Z_{ij11} + Z_{ij21} + Z_{ij31} + Z_{ij41}\right)/\left(V'_{ij11} + V'_{ij21} + V'_{ij31} + V'_{ij41}\right)$ that is averaged over complete simulated runs to provide $\tilde{\theta}_{ij}$ and used to determine $\text{se}\left(\tilde{\theta}_{ij}\right)$. The reason for this change is pragmatic: without it estimates show excessive bias and standard errors are too small or sometimes non-existent as equation (2) involves the square root of a negative value. Use of $V'_{ijc1}$ largely avoids these problem, as $E(Z_{ijc1})$ is closer to $\theta V'_{ijc1}$ than it is to $\theta V_{ijc1}$ and $\text{var}(Z_{ijc1})$ is closer to $V'_{ijc1}$ than it is to $V_{ijc1}$. In the unstratified case the sample sizes per treatment at the first interim analysis are quite large, and so this level of attention to detail is unnecessary. In the stratified case, it is the sample sizes within centre that determine the accuracy of the procedure, and without the use of (7) these are now too small to guarantee the accuracy of the estimates, or the existence of the standard errors.

Table 7 compares a naïve analysis in which pairs of treatments are compared using the data available at the last interim analysis in which both were present but ignoring the sequential nature of the trial, with the RB2 method described above. The number of reverse simulations was set at 10 million. It can be seen that the effect of allowance for the sequential design is to reduce the magnitude of the estimates of the advantage of $T_1$ over each of $T_2$ and $T_4$, while the estimate of the advantage of $T_1$ over $T_3$ is hardly changed. The corresponding confidence intervals are all widened. The other estimates of treatment effects have also been reduced in magnitude, but the effect on their standard errors is less marked. The SAS programs leading to the RB2 analyses shown in Table 7 is provided as supplementary material of this paper.

Table 8 shows the results from 1000 replicate simulations of a situation in which $T_1$ is the best treatment. To achieve a feasible computational time, one million reverse simulations are used in each analysis. Furthermore, for ease of computation, Option 1 is chosen so that a single set of reverse simulations will yield estimates and confidence intervals for all treatment comparisons. For comparison, the results from naïve analyses based on the test statistics Z and V comparing the final samples simulated from each treatment (that is using Option 1) are also shown. The comparisons of $T_1$ with the three rival treatments each lead to overestimation of treatment effect when the naïve analysis is used, whereas the estimates drawn from RB2 show very little bias. The results for the other comparisons are similar for the two approaches, with RB2 being a little less biased. In most of the simulated realisations, the timing of these comparisons will have been determined by the completion of others, and so the effects of the sequential design would be expected to be less marked. The coverage probabilities for confidence intervals based on the naïve approach are too low, while those for RB2 are satisfactory. Other simulations were conducted in which each RB2 analysis depended on only 100,000 reverse simulations. These led to less accurate estimation and markedly conservative confidence intervals. It appears that, provided that sufficient reverse simulations are used, RB2 leads to accurate analyses that overcome the potential bias inherent in the use of data-dependent elimination and stopping rules.

## 6. Conclusions

The approach presented here for estimation following a sequential trial is quite general, and can be implemented for a wide variety of designs. In the case of a comparison of a single experimental treatment with a single control arm, the method works and provides satisfactory results, as has been demonstrated in Section 4 above. However, there are already numerous methods of computing point and interval estimates in the two-treatment context. In particular, methods based on orderings of the final sample space are just as good for computing point estimates and more accurate for finding confidence intervals than the approach introduced here. They are also less computationally demanding.

The utility of the approach described here is in more complicated designs comparing multiple treatments or with flexible adaptive features, as reverse simulation is based only on the form of the stopping rules implemented and not on their theoretical properties. The method has been illustrated and evaluated for one particular form of comparison of four treatments which motivated its development, but its implementation is certainly not restricted to that design.

The claim for the unbiasedness of estimates produced using Method RB2 is underpinned by rigorous asymptotic theory, and the simulation results obtained for their accuracy in Section 5 are satisfactory. The method for deriving confidence intervals is less secure as it depends on two unverified assumptions: that the expected conditional variance of the unbiased estimate at the first interim analysis can be approximated by its observed value from reverse simulations, and that the adjusted estimate follows the normal distribution. Simulations in Sections 4 and 5 demonstrate that the resulting intervals are conservative but serviceable. It should be repeated that the number of reverse simulations needed to achieve satisfactory results is large. Here, in single demonstration analyses, 10 million replicates were used. In earlier work, we found that using fewer replicates led to less satisfactory results.

**Acknowledgement**

The first author is grateful for discussions with Amalia Magaret and Shevin Jacob concerning the design of the sepsis trial described in Section 2. This work is independent research arising in part from Dr. Jaki's Senior Research Fellowship (NIHR-SRF-2015-08-001) supported by the National Institute for Health Research. Funding for this work was also provided by the Medical Research Council (MR/M005755/1). The views expressed in this publication are those of the authors and not necessarily those of the NHS, the National Institute for Health Research, or the Department of Health.

**Data availability statement**

All of the data that support the findings of this study are available within the paper itself.

**Table 1: Properties of the four treatment design from million-fold simulations**

*$win_1$ = proportion of runs in which $T_1$ wins*
*$elim_4$ = proportion of runs in which $T_4$ is eliminated*
*nod = proportion of runs in which: for Cases 1-8 and Mixed Cases I –II, $T_1$ and $T_2$ are declared no different from one another; for Cases 9-12 and Mixed Case III, $T_1$, $T_2$ and $T_3$ are declared no different from one another; for Cases 13-16 and Mixed Case IV, all treatments are declared no different from one another*
*still = proportion of runs in which not all treatment comparisons are resolved after 2772 responses*

| Case | $p_1$ | $p_2$ | $p_3$ | $p_4$ | E(n) | $win_1$ | $elim_4$ | nod | still |
|---|---|---|---|---|---|---|---|---|---|
| 1 | 0.500 | 0.400 | 0.400 | 0.400 | 1426 | 0.819 | 0.920 | 0.045 | 0.000 |
| 2 | 0.600 | 0.500 | 0.500 | 0.500 | 1427 | 0.819 | 0.920 | 0.044 | 0.000 |
| 3 | 0.692 | 0.600 | 0.600 | 0.600 | 1537 | 0.816 | 0.916 | 0.043 | 0.004 |
| 4 | 0.771 | 0.692 | 0.692 | 0.692 | 1765 | 0.802 | 0.902 | 0.039 | 0.039 |
| Mixed Case I (Cases 1-4) | | | | | 1531 | 0.819 | 0.918 | 0.043 | 0.004 |
| 5 | 0.500 | 0.500 | 0.400 | 0.400 | 1389 | 0.025 | 0.975 | 0.901 | 0.000 |
| 6 | 0.600 | 0.600 | 0.500 | 0.500 | 1411 | 0.025 | 0.975 | 0.903 | 0.000 |
| 7 | 0.692 | 0.692 | 0.600 | 0.600 | 1540 | 0.026 | 0.974 | 0.901 | 0.002 |
| 8 | 0.771 | 0.771 | 0.692 | 0.692 | 1803 | 0.026 | 0.966 | 0.885 | 0.024 |
| Mixed Case II (Cases 5-8) | | | | | 1524 | 0.026 | 0.975 | 0.975 | 0.001 |
| 9 | 0.500 | 0.500 | 0.500 | 0.400 | 1540 | 0.005 | 0.988 | 0.861 | 0.000 |
| 10 | 0.600 | 0.600 | 0.600 | 0.500 | 1583 | 0.005 | 0.988 | 0.861 | 0.000 |
| 11 | 0.692 | 0.692 | 0.692 | 0.600 | 1752 | 0.005 | 0.987 | 0.857 | 0.003 |
| 12 | 0.771 | 0.771 | 0.771 | 0.692 | 2066 | 0.005 | 0.975 | 0.814 | 0.057 |
| Mixed Case III (Cases 9-12) | | | | | 1722 | 0.005 | 0.987 | 0.857 | 0.003 |
| 13 | 0.500 | 0.500 | 0.500 | 0.500 | 1795 | 0.002 | 0.066 | 0.785 | 0.001 |
| 14 | 0.600 | 0.600 | 0.600 | 0.600 | 1862 | 0.002 | 0.066 | 0.782 | 0.004 |
| 15 | 0.692 | 0.692 | 0.692 | 0.692 | 2071 | 0.002 | 0.066 | 0.748 | 0.053 |
| 16 | 0.771 | 0.771 | 0.771 | 0.771 | 2381 | 0.001 | 0.064 | 0.591 | 0.266 |
| Mixed Case IV (Cases 13-16) | | | | | 2028 | 0.002 | 0.066 | 0.760 | 0.036 |

**Table 2: Details of 12 realisations of the triangular design and of two simple forms of analysis**

*Terminal values of the number of interim analyses, total sample size, the numbers of successes on $T_1$ and $T_2$ and of the statistics Z and V are shown as int*, n*, $S_1$*, $S_2$*, Z* and V* respectively. Patients are evenly divided between the two treatments so that $n_1^* = n_2^* = ½n^*$.*

*b* denotes the boundary crossed, with 0 denoting the lower boundary and 1 the upper boundary.*

*For the naïve analysis, the estimated value of $\theta$ is Z*/V* with 95% confidence interval $(\theta_L, \theta_U) = (\hat{\theta} \pm 1.96/\sqrt{V^*})$.*

*The orderings analysis is based on the ordering of Fairbanks and Madsen [21] and computed following [19, 20].*

| Case | Terminal data | | | | | | | Naïve analysis | | | | Orderings analysis | | | |
|---|---|---|---|---|---|---|---|---|---|---|---|---|---|---|---|
| | int* | n* | $S_1$* | $S_2$* | Z* | V* | b* | p-val | $\hat{\theta}$ | $\theta_L$ | $\theta_U$ | p-val | $\theta_M$ | $\theta_L$ | $\theta_U$ |
| 1 | 2 | 144 | 35 | 59 | –12.0 | 8.160 | 0 | 1.000 | –1.471 | –2.157 | –0.784 | 1.000 | –1.470 | –2.156 | –0.783 |
| 2 | 3 | 216 | 68 | 87 | –9.5 | 10.943 | 0 | 0.998 | –0.868 | –1.461 | –0.276 | 0.997 | –0.857 | –1.454 | –0.256 |
| 3 | 4 | 288 | 102 | 118 | –8.0 | 12.986 | 0 | 0.987 | –0.616 | –1.160 | –0.072 | 0.983 | –0.599 | –1.149 | –0.044 |
| 4 | 10 | 720 | 284 | 285 | –0.5 | 29.833 | 0 | 0.537 | –0.017 | –0.376 | 0.342 | 0.485 | 0.007 | –0.358 | 0.378 |
| 5 | 8 | 576 | 201 | 201 | 0.0 | 30.359 | 0 | 0.500 | 0.000 | –0.356 | 0.356 | 0.464 | 0.017 | –0.344 | 0.382 |
| 6 | 13 | 936 | 275 | 259 | 8.0 | 57.337 | 0 | 0.144 | 0.140 | –0.119 | 0.398 | 0.089 | 0.187 | –0.084 | 0.468 |
| 7 | 9 | 648 | 252 | 222 | 15.0 | 31.819 | 1 | 0.004 | 0.471 | 0.124 | 0.819 | 0.007 | 0.454 | 0.097 | 0.807 |
| 8 | 6 | 432 | 120 | 88 | 16.0 | 26.963 | 1 | 0.001 | 0.593 | 0.216 | 0.971 | 0.003 | 0.563 | 0.168 | 0.949 |
| 9 | 6 | 432 | 161 | 130 | 15.5 | 23.745 | 1 | 0.001 | 0.653 | 0.251 | 1.055 | 0.002 | 0.623 | 0.205 | 1.034 |
| 10 | 5 | 360 | 135 | 108 | 13.5 | 19.744 | 1 | 0.001 | 0.684 | 0.243 | 1.125 | 0.002 | 0.676 | 0.231 | 1.120 |
| 11 | 5 | 360 | 124 | 92 | 16.0 | 21.600 | 1 | 0.000 | 0.741 | 0.319 | 1.162 | 0.001 | 0.704 | 0.260 | 1.137 |
| 12 | 3 | 216 | 82 | 55 | 13.5 | 12.527 | 1 | 0.000 | 1.078 | 0.524 | 1.631 | 0.000 | 1.075 | 0.519 | 1.629 |

**Table 3: Analyses of the 12 realisations of the triangular design based on Rao-Blackwellisation**

| Case | Method RB1 | | | | Method RB2 | | | | |
|---|---|---|---|---|---|---|---|---|---|
| | $\tilde{\theta}$ | se | $\theta_L$ | $\theta_U$ | % complete | $\tilde{\theta}$ | se | $\theta_L$ | $\theta_U$ |
| 1 | −1.463 | 0.360 | −2.169 | −0.757 | 99.3 | −1.473 | 0.383 | −2.225 | −0.722 |
| 2 | −0.823 | 0.325 | −1.461 | −0.185 | 89.3 | −0.834 | 0.334 | −1.488 | −0.180 |
| 3 | −0.560 | 0.298 | −1.145 | 0.025 | 79.9 | −0.567 | 0.295 | −1.145 | 0.010 |
| 4 | 0.046 | 0.204 | −0.354 | 0.447 | 55.7 | 0.046 | 0.158 | −0.263 | 0.356 |
| 5 | 0.051 | 0.201 | −0.342 | 0.445 | 67.0 | 0.052 | 0.183 | −0.307 | 0.411 |
| 6 | 0.224 | 0.166 | −0.101 | 0.549 | 17.0 | 0.227 | 0.158 | −0.081 | 0.536 |
| 7 | 0.420 | 0.197 | 0.033 | 0.806 | 63.7 | 0.424 | 0.185 | 0.062 | 0.787 |
| 8 | 0.519 | 0.214 | 0.100 | 0.939 | 56.0 | 0.529 | 0.213 | 0.110 | 0.947 |
| 9 | 0.580 | 0.226 | 0.136 | 1.024 | 54.9 | 0.584 | 0.229 | 0.135 | 1.033 |
| 10 | 0.653 | 0.239 | 0.184 | 1.122 | 85.7 | 0.658 | 0.245 | 0.179 | 1.138 |
| 11 | 0.655 | 0.238 | 0.188 | 1.122 | 58.5 | 0.671 | 0.243 | 0.195 | 1.147 |
| 12 | 1.059 | 0.291 | 0.490 | 1.629 | 95.8 | 1.069 | 0.312 | 0.457 | 1.680 |

**Table 4: Evaluation of the naïve and the Rao-Blackwellisation methods based on 1,000-fold simulations**

| | Naïve | | | Method RB1 | | | Method RB2 | | |
|---|---|---|---|---|---|---|---|---|---|
| **True value of $\theta$** | **0** | **0.246** | **0.405** | **0** | **0.246** | **0.405** | **0** | **0.246** | **0.405** |
| Estimate of $\theta$ | –0.069 | 0.244 | 0.459 | –0.001 | 0.248 | 0.410 | –0.006 | 0.246 | 0.408 |
| Standard deviation | 0.209 | 0.227 | 0.213 | 0.213 | 0.182 | 0.203 | 0.233 | 0.187 | 0.196 |
| Standard error | 0.184 | 0.154 | 0.169 | 0.209 | 0.184 | 0.197 | 0.201 | 0.175 | 0.190 |
| $\theta_L$ | –0.430 | –0.058 | 0.128 | –0.408 | –0.113 | 0.025 | –0.399 | –0.096 | 0.034 |
| $\theta_U$ | 0.293 | 0.546 | 0.790 | 0.410 | 0.609 | 0.795 | 0.388 | 0.589 | 0.781 |
| Probability that $\theta \in (\theta_L, \theta_U)$ | 0.943 | 0.932 | 0.920 | 0.976 | 0.976 | 0.972 | 0.958 | 0.967 | 0.971 |

**Table 5: Raw data from a single simulation of the four treatment design**

| Treatment | Interim | Centre | n | S | Sample size at each interim | Number of successes at each interim |
|---|---|---|---|---|---|---|
| 1 | 12 | 1 | 103 | 83 | 11, 18, 30, 41, 50, 57, 65, 76, 86, 92, 98, 103 | 10, 17, 27, 35, 41, 46, 53, 63, 69, 74, 78, 83 |
| | | 2 | 100 | 67 | 10, 16, 25, 33, 41, 49, 60, 71, 82, 88, 96, 100 | 10, 14, 20, 25, 30, 34, 40, 47, 58, 61, 65, 67 |
| | | 3 | 104 | 64 | 7, 17, 25, 35, 44, 55, 63, 68, 72, 83, 90, 104 | 6, 11, 16, 20, 26, 32, 36, 41, 43, 49, 55, 64 |
| | | 4 | 125 | 68 | 8, 21, 28, 35, 45, 55, 64, 73, 84, 97, 112, 125 | 4, 13, 15, 20, 27, 34, 38, 45, 48, 53, 62, 68 |
| | | **Total** | **432** | **282** | | |
| 2 | 4 | 1 | 39 | 25 | 12, 24, 31, 39 | 9, 17, 19, 25 |
| | | 2 | 30 | 13 | 6, 13, 25, 30 | 4, 8, 12, 13 |
| | | 3 | 35 | 21 | 7, 16, 22, 35 | 5, 11, 15, 21 |
| | | 4 | 40 | 11 | 11, 19, 30, 40 | 1, 5, 8, 11 |
| | | **Total** | **144** | **70** | | |
| 3 | 12 | 1 | 111 | 85 | 9, 19, 29, 39, 48, 57, 67, 74, 85, 91, 102, 111 | 8, 15, 21, 27, 33, 41, 49, 56, 65, 70, 79, 85 |
| | | 2 | 94 | 56 | 7, 15, 24, 32, 40, 49, 57, 64, 72, 79, 88, 94 | 5, 9, 15, 22, 28, 31, 33, 38, 44, 47, 52, 56 |
| | | 3 | 111 | 60 | 9, 17, 25, 32, 42, 50, 58, 68, 76, 90, 101, 111 | 3, 5, 8, 13, 21, 27, 31, 37, 41, 48, 55, 60 |
| | | 4 | 116 | 45 | 11, 21, 30, 41, 50, 60, 70, 82, 91, 100, 105, 116 | 4, 7, 12, 15, 18, 23, 26, 34, 37, 42, 44, 45 |
| | | **Total** | **432** | **246** | | |
| 4 | 5 | 1 | 50 | 32 | 9, 15, 23, 36, 50 | 5, 11, 17, 24, 32 |
| | | 2 | 47 | 27 | 9, 20, 32, 42, 47 | 6, 11, 16, 24, 27 |
| | | 3 | 40 | 18 | 11, 19, 28, 32, 40 | 5, 8, 12, 14, 18 |
| | | 4 | 43 | 16 | 7, 18, 25, 34, 43 | 3, 9, 10, 13, 16 |
| | | **Total** | **180** | **93** | | |

**Table 6: Comparative data derived from Table 5**

| Comparison | Interim | Site | Z | V | $\hat{\theta}$ | Conclusion |
|---|---|---|---|---|---|---|
| $T_1$ vs $T_2$ | 4 | 1 | 4.25 | 3.75 | 1.133 | $T_1$ knocks out $T_2$ at 4th interim |
| | | 2 | 5.10 | 3.76 | 1.356 | |
| | | 3 | –0.50 | 4.25 | –0.118 | |
| | | 4 | 5.53 | 4.53 | 1.221 | |
| | | **Total** | **14.38** | **16.28** | **0.883** | |
| $T_1$ vs $T_3$ | 12 | 1 | 2.14 | 9.02 | 0.237 | $T_1$ knocks out $T_3$ at 12th interim |
| | | 2 | 3.60 | 11.24 | 0.320 | |
| | | 3 | 4.02 | 13.11 | 0.307 | |
| | | 4 | 9.39 | 14.98 | 0.627 | |
| | | **Total** | **19.15** | **48.35** | **0.396** | |
| $T_1$ vs $T_4$ | 5 | 1 | 4.50 | 4.93 | 0.913 | $T_1$ knocks out $T_4$ at 5th interim |
| | | 2 | 3.44 | 5.00 | 0.688 | |
| | | 3 | 2.95 | 5.23 | 0.564 | |
| | | 4 | 5.01 | 5.49 | 0.912 | |
| | | **Total** | **15.91** | **20.64** | **0.771** | |
| $T_2$ vs $T_3$ | 4 | 1 | –1.00 | 4.33 | –0.231 | No conclusion |
| | | 2 | –3.94 | 3.81 | –1.034 | |
| | | 3 | 3.23 | 4.18 | 0.773 | |
| | | 4 | –1.84 | 4.41 | –0.417 | |
| | | **Total** | **–3.54** | **16.73** | **–0.212** | |
| $T_2$ vs $T_4$ | 4 | 1 | –0.48 | 4.24 | –0.113 | No conclusion |
| | | 2 | –2.42 | 4.37 | –0.554 | |
| | | 3 | 2.72 | 4.17 | 0.652 | |
| | | 4 | –1.97 | 4.03 | –0.489 | |
| | | **Total** | **–2.15** | **16.81** | **–0.128** | |
| $T_3$ vs $T_4$ | 5 | 1 | 1.16 | 5.47 | 0.212 | No conclusion |
| | | 2 | 2.71 | 5.02 | 0.540 | |
| | | 3 | 1.02 | 5.11 | 0.200 | |
| | | 4 | –0.28 | 5.36 | –0.052 | |
| | | **Total** | **4.62** | **20.97** | **0.220** | |

**Table 7: Analyses of the data from the single simulated run of the sequential four treatment comparison shown in Tables 5 and 6**

*In the naïve analyses, the sequential nature of the trial is ignored*

*The Rao-Blackwellisation method, RB2, is based on 10 million replicate reverse simulations*

| Comparison | Naïve | | | | Proportion complete | RB2 | | | |
|---|---|---|---|---|---|---|---|---|---|
| | $\hat{\theta}$ | se | $\theta_L$ | $\theta_U$ | | $\tilde{\theta}$ | se | $\theta_L$ | $\theta_U$ |
| $T_1$ vs $T_2$ | 0.883 | 0.248 | 0.347 | 1.319 | 0.7381 | 0.869 | 0.286 | 0.309 | 1.429 |
| $T_1$ vs $T_3$ | 0.396 | 0.144 | 0.114 | 0.678 | 0.0199 | 0.405 | 0.220 | –0.027 | 0.837 |
| $T_1$ vs $T_4$ | 0.771 | 0.220 | 0.340 | 1.202 | 0.3050 | 0.667 | 0.256 | 0.165 | 1.169 |
| $T_2$ vs $T_3$ | –0.212 | 0.244 | –0.690 | 0.266 | 0.7381 | –0.167 | 0.255 | –0.667 | 0.333 |
| $T_2$ vs $T_4$ | –0.128 | 0.244 | –0.606 | 0.350 | 0.7381 | –0.069 | 0.249 | –0.557 | 0.418 |
| $T_3$ vs $T_4$ | 0.220 | 0.218 | –0.207 | 0.647 | 0.3050 | 0.165 | 0.225 | –0.277 | 0.606 |

**Table 8: Evaluation of the naïve method and the Rao-Blackwellisation method RB2 in the four treatment case**

*Both evaluations are based on 1,000-fold simulations and each RB2 analysis employed 1,000,000 reverse-simulations*

*The RB2 results are based on the 893 replicates in which 1000 or more reverse simulations were complete.*

| **Method** | **Naïve** | | | | | | **RB2** | | | | | |
|---|---|---|---|---|---|---|---|---|---|---|---|---|
| **Comparison** | **$T_1$ vs $T_2$** | **$T_1$ vs $T_3$** | **$T_1$ vs $T_4$** | **$T_2$ vs $T_3$** | **$T_2$ vs $T_4$** | **$T_3$ vs $T_4$** | **$T_1$ vs $T_2$** | **$T_1$ vs $T_3$** | **$T_1$ vs $T_4$** | **$T_2$ vs $T_3$** | **$T_2$ vs $T_4$** | **$T_3$ vs $T_4$** |
| **True value of $\theta$** | **0.693** | **0.405** | **0.405** | **–0.288** | **–0.288** | **0.000** | **0.693** | **0.405** | **0.405** | **–0.288** | **–0.288** | **0.000** |
| Estimate of $\theta$ | 0.771 | 0.462 | 0.461 | –0.302 | –0.304 | –0.002 | 0.690 | 0.405 | 0.395 | –0.291 | –0.301 | –0.010 |
| Standard deviation | 0.217 | 0.183 | 0.185 | 0.208 | 0.213 | 0.178 | 0.242 | 0.202 | 0.204 | 0.225 | 0.226 | 0.182 |
| Standard error | 0.193 | 0.160 | 0.160 | 0.196 | 0.196 | 0.165 | 0.252 | 0.219 | 0.219 | 0.213 | 0.213 | 0.185 |
| $\theta_L$ | 0.393 | 0.149 | 0.147 | –0.687 | –0.688 | –0.325 | 0.195 | –0.024 | –0.035 | –0.708 | –0.719 | –0.372 |
| $\theta_U$ | 1.149 | 0.775 | 0.773 | 0.083 | 0.081 | 0.322 | 1.184 | 0.833 | 0.824 | 0.126 | 0.117 | 0.352 |
| Probability that $\theta \in (\theta_L, \theta_U)$ | 0.932 | 0.925 | 0.924 | 0.941 | 0.935 | 0.943 | 0.951 | 0.973 | 0.965 | 0.962 | 0.955 | 0.968 |

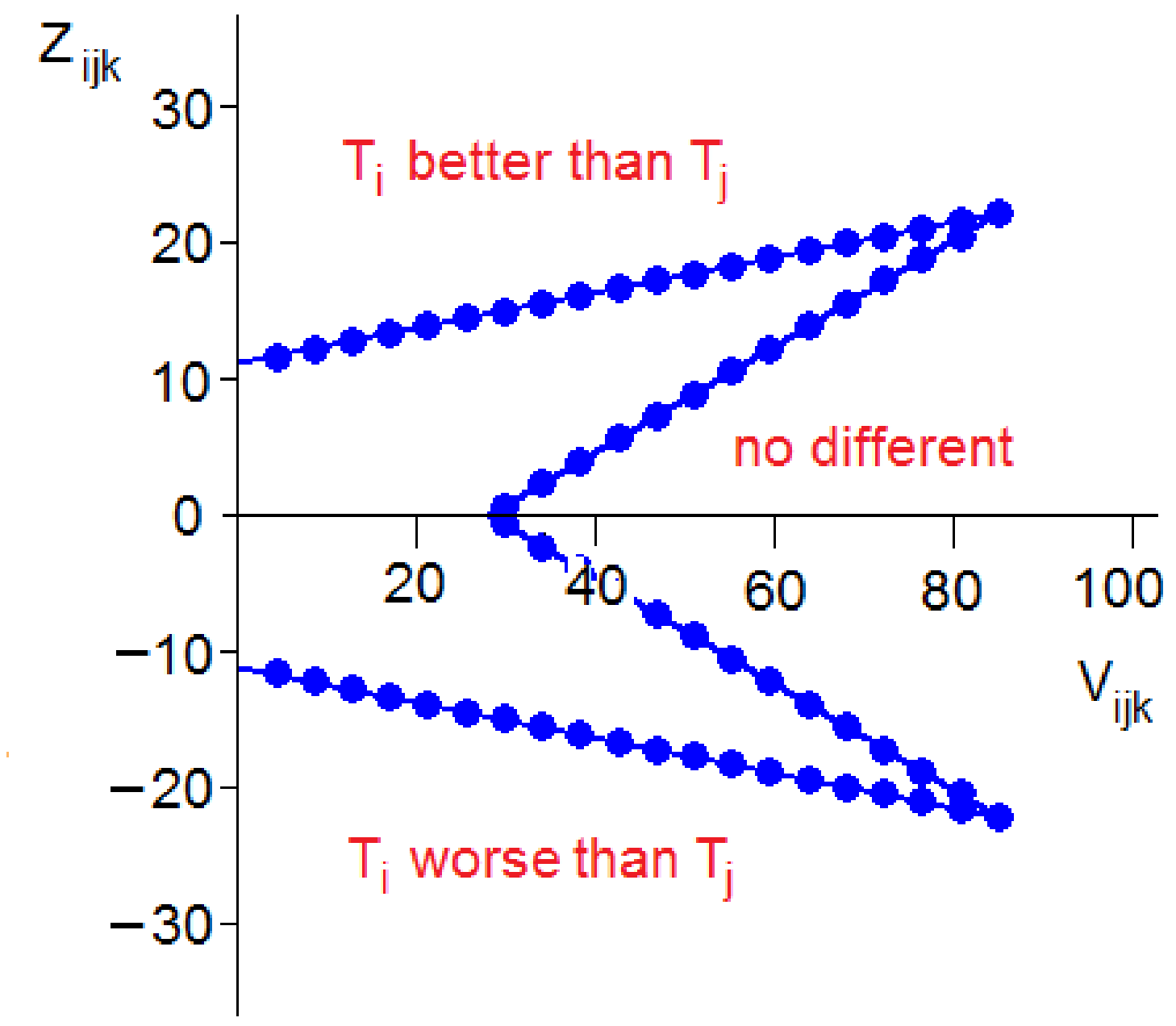

*Figure 1: The elimination and stopping rule for a single pair of treatments*

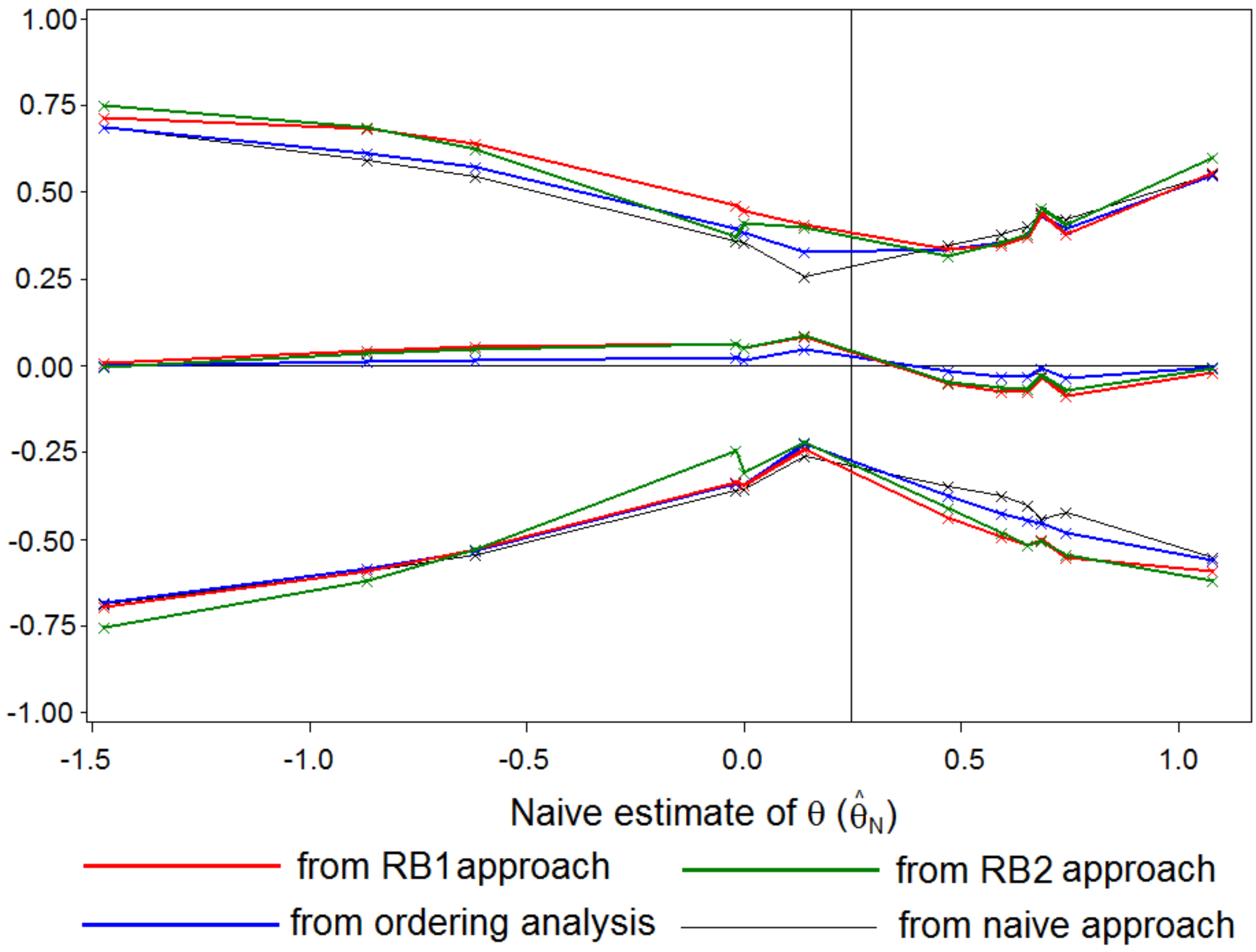

*Figure 2: Estimates and 95% confidence limits for $\theta$ from the Rao-Blackwellisation approaches, the orderings analysis and the naïve approach - with the naïve estimate subtracted - plotted against the naïve estimate for Cases 1-12*